\documentclass{ws-procs9x6}

\begin{document}

\title{Theoretical aspects of ultra high energy cosmic rays }

\author{Pasquale Blasi}

\address{INAF/Osservatorio Astrofisico di Arcetri\\ 
Largo E. Fermi, 5 - I50125 Firenze, ITALY
E-mail: blasi@arcetri.astro.it}

\maketitle

\abstracts{
We review the basic ideas on the origin of cosmic rays with 
energy in excess of $\sim 10^{19}$ eV, in the light of the most recent 
observational findings. The limited statistics of events detected by the two 
largest experiments currently operating does not allow as yet to claim the 
detection of the GZK feature in the cosmic ray spectrum or the lack of it. 
Although extragalactic point sources seem to be preferred on the basis of 
the small scale anisotropies detected by AGASA, the possibility that UHECRs 
with energy in excess of $10^{20}$ eV may be the result of the decay of 
supermassive particles is not ruled out by present data. It seems clear that 
if we want to have a clear idea of the origin of UHECRs, we have to auspicate 
the realization of larger, better experiments, such as the ongoing Auger 
project and the space-borne EUSO experiment.
}

\section{Introduction}

Ninety years after the discovery of cosmic rays, their origin is still
surrounded by many questions still seeking an answer. May be the most 
resilient of these puzzles is the origin of the highest energy end of
the cosmic ray spectrum, that lies now near $10^{20}$ eV. The search 
for the sources of cosmic rays at ZeV energies is made harder by two
factors: first, it is difficult to envision astrophysical accelerators 
able to achieve such large energies \cite{olinto}, and second, the laws 
of physics predict a flux suppression at $\sim 5\times 10^{19}$ eV, that 
has not yet been clearly detected. This suppression, known as the GZK feature 
\cite{gzk}, is expected to appear when cosmic ray protons start to feel 
the inelastic collisions with the cosmic microwave background (photopion 
production), a process that has a threshold at $\sim 5\times 10^{19}$ eV.
While the first point is mainly a matter of theoretical investigation, 
the second issue is mainly an observational issue. 
The status of observations has been summarized in this conference 
\cite{nagano}, and we refer the reader to that contribution for details, 
while we concentrate here on the possible interpretations of the observations 
in terms of origin and propagation of UHECRs. 

The paper is organized as follows: in \S \ref{sec:gzk} we address the
issue of the statistical significance of present observations with AGASA 
and HiRes; in \S \ref{sec:astro} we summarize some plausible astrophysical 
sources of UHECRs; in \S \ref{sec:td} we describe the predictions and status 
of the art of top-down models of UHECR production. We conclude in 
\S \ref{sec:concl}.

\section{First shadows of the GZK feature?}\label{sec:gzk}

\begin{figure}[ht]
\centerline{\epsfxsize=2.6in\epsfbox{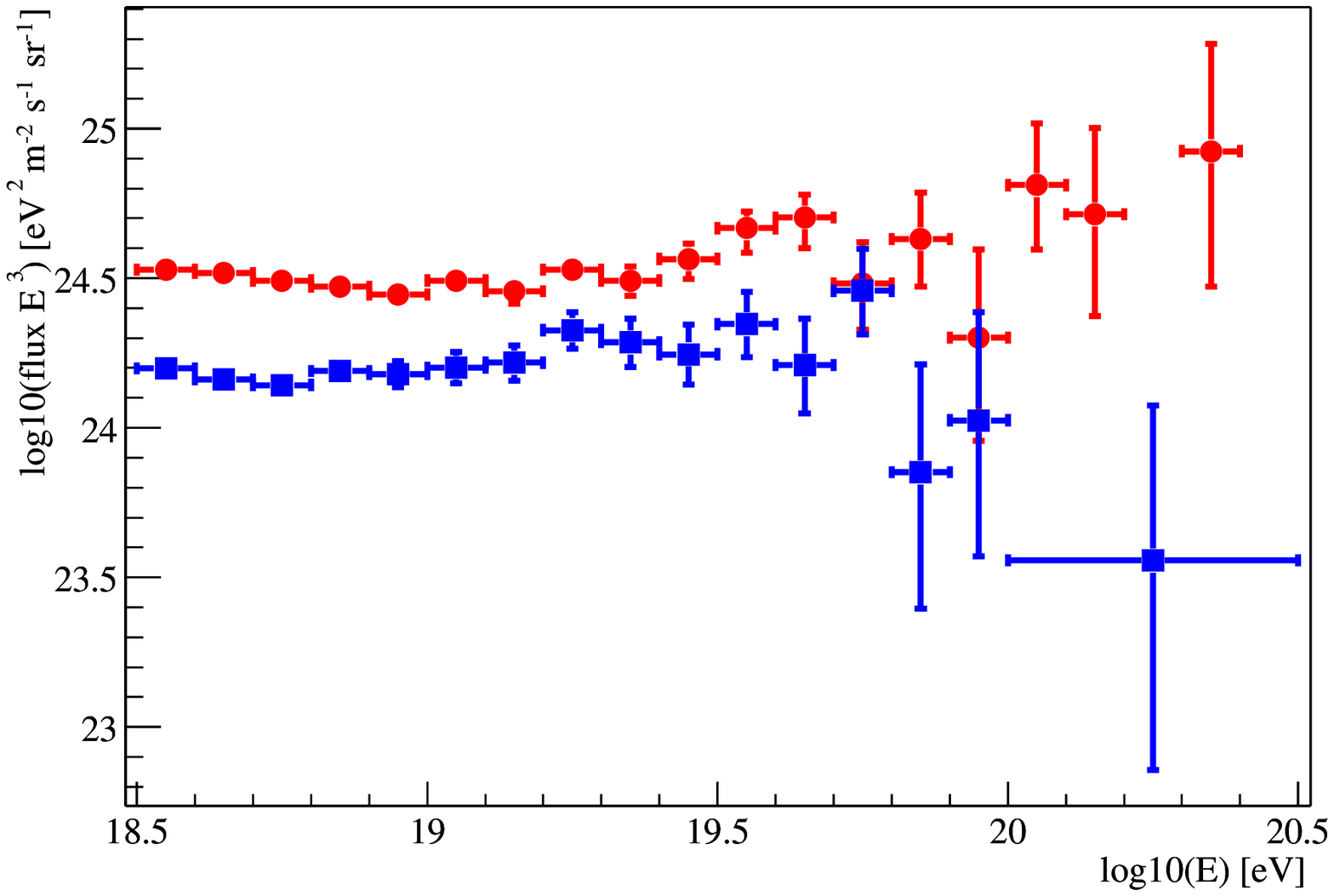}
\epsfxsize=2.6in\epsfbox{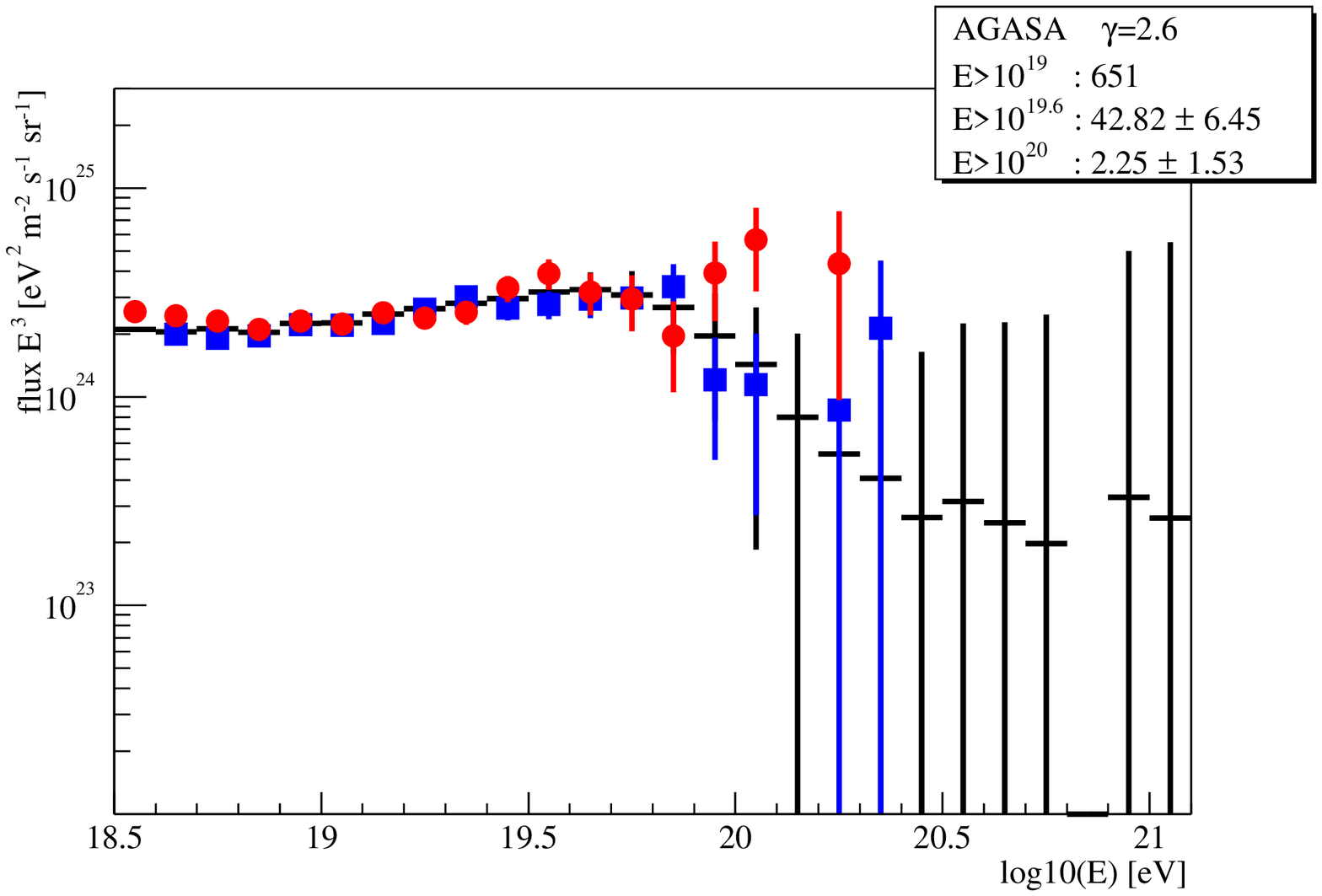}}
\caption{{\it left}) AGASA data (circles) and HiResI data (squares); 
{\it right}) Results of the simulation for a $15\%$ systematic 
error in the energy determination of AGASA $^4$.}
\label{agasahires}
\end{figure}

As summarized in Ref. \cite{nagano}, current observations by AGASA and
HiRes show some discrepancy in both the absolute flux, and the spectral
shape of the UHECRs. In particular the HiRes experiment seems to show some
evidence for the presence of the expected GZK feature. The AGASA data
show no sign of the GZK suppression, and are consistent with a power
law extrapolation of the lower energy spectrum. Despite some claims of
detection of the GZK feature \cite{bw}, the investigation reported in 
\cite{us} suggests that the situation may in fact be more subtle and
that more data are needed to substantiate such claims. As illustrated 
in Fig. \ref{agasahires} (left panel), the two data sets have an offset 
that might be accomodated if a systematic error in the energy determination 
of $\sim 15\%$ in each experiment were present. In Ref. \cite{us} 
the propagation of UHECRs has been followed through a numerical simulation, 
so to keep track of the uncertainties connected with the limited statistics 
of events of each experiment. At energies larger than $10^{20}$ eV, the 
discrepancy between the two experiments has been found to be at the level
of $2.6\sigma$ without any correction for systematics. If a systematic
error in the energy determination of $\sim 15\%$ is assumed in each 
experiment, the discrepancy is reduced further. It is interesting to note 
that this systematic error would be perfectly compatible with the estimates 
of systematic errors published independently by the two collaborations 
\cite{astro0209422,HIRES2}.   
Using a Montecarlo simulation, it is possible to evaluate the uncertainty
in the predicted fluxes, as due to the intrinsic stochasticity of the 
photopion production process and to cosmic variance. If these uncertainties
are taken into account, the discrepancy between the two data sets is at the
$1.8\sigma$ level without correcting for the systematics, and at the 
$1.5\sigma$ level if such correction is taken into account (right panel in 
Fig. \ref{agasahires}\cite{us}) . It is possible that the first shadows of 
the GZK feature are appearing but any claim either of detection of the GZK 
feature or lack of it is justified only at the $\sim 2\sigma$ level, 
and therefore far from being conclusive. 

\section{Astrophysical sources of UHECRs}\label{sec:astro}

Astrophysical sources distributed homogeneously in the universe, or 
following the observed density field \cite{bbo} produce a spectrum
of UHECRs with a GZK feature, more pronounced in general for steeper 
injection spectra \cite{grigo88}. Therefore the firm detection of this feature
in the spectrum may serve as the most convincing evidence for an extragalactic
astrophysical origin of UHECRs. Here we briefly summarize the potential 
astrophysical sources of UHECRs and the acceleration processes that are 
supposed to be at work. 

\subsection{Active and dead Galactic Nuclei}

The term Active Galactic Nucleus (AGN) is used here to identify a very 
broad class of objects, characterized by the presence of a central massive 
black hole, and fueled by an accretion flow. The presence of jets is 
probably a characteristic of many AGNs. When the fuel is almost 
exhausted, the AGN stops being active, leaving behind a {\it dead quasar} 
\cite{boldt}. Cosmic rays can be accelerated in the central regions of AGNs, 
in the jets of some special class of AGNs and in the accretion disks of dead 
quasars.
In the central regions of AGNs energy losses and diffusive confinement play 
together to limit the maximum energy to $10^6-10^7$ GeV \cite{norman}. 
Hot spots in F-R II radio galaxies appear to be more promising sites for 
the generation of UHECRs, due to the lower photon density and lower values 
of the magnetic field in the acceleration region \cite{rachenB93}. The hot 
spot is terminated by a strong shock at which particles may be accelerated 
diffusively reaching a maximum energy that can be as high as $10^{21}$ eV 
or more. The spectrum of UHECRs at the Earth is expected to have a pronounced
GZK cutoff, because the closest of these sources is at redshift of $\sim 
0.3$. The propagation of UHECRs from generic active galactic nuclei has been 
recently investigated by \cite{bereagn}.

Recently it has been proposed \cite{boldt} that unipolar induction in 
dead quasars may energize charged particles to ultra high energies: in these 
objects energy losses due to photopion production and curvature radiation are 
expected to be less important than for active galaxies. Although dead quasars 
are not expected to be bright sources, it has been proposed \cite{levinson} 
that they may show gamma ray emission, due to curvature radiation of UHECRs 
during the acceleration process. Unfortunately, at present no detailed 
calculation of the spectrum of UHECRs generated by dead quasars exists in the 
literature, so that it is difficult to predict in a more quantitative way the 
spectrum of particles detected at the Earth.

Other astrophysical objects may be suitable candidates as sources of 
UHECRs, although a specific model may not have been worked out in the details. 
Recently, evidence has been found for a correlation of the arrival directions 
of cosmic rays with energy above $2\times 10^{19}$ eV with the spatial location
of BL Lac objects \cite{tkachev} and a possible correlation with EGRET 
sources \cite{gamma}. Many of the BL Lacs are at large redshifts. 
It is worth stressing however that this correlation concerns only
cosmic rays with energy below $\sim 5-6\times 10^{19}$ eV. At these 
energies the pathlength is large enough to allow for the arrival of particles 
from sources at redshift in excess of 0.1. Therefore the correlation does 
not imply any need for new physics, and does not suggest any plausible
explanation for the events with energy larger than $5\times 10^{19}$ eV,
whose sources remain unidentified.

\subsection{Neutron Stars}

Rapidly rotating strongly magnetized neutron stars are very efficient unipolar 
inductors, with an electromotive force ({\it emf}) that may reach $10^{21}$ V 
\cite{venka}, although this potential is likely to be partially short-cut by 
the electron-positron pairs in the magnetosphere of the neutron star.
The {\it emf} is available in gap regions where the 
condition $\vec E \cdot \vec B = 0$ is violated \cite{ruderman}. The maximum 
achievable energy is typically around $10^{15}$ eV \cite{bible}, mainly due to 
the limitation imposed by curvature radiation energy losses.
In \cite{boe} a phenomenological approach was adopted to rule in favor of 
acceleration of heavy nuclei to extremely high energies at the light cylinder 
of young neutron stars: it is well known that most energy lost by a spinning
neutron star is not in the form of radiation \cite{coroniti}, but is rather
converted into kinetic energy of a relativistic wind. For a Crab-like pulsar, 
the Lorentz factor of the wind must be of the order of $\Gamma_W\sim 10^6-10^7$
\cite{coroniti}, although it is not clear how this relativistic motion 
is achieved. If at the light cylinder a small amount of Iron 
nuclei or protons is present, the nuclei may acquire the same Lorentz factor 
of the wind, that for young neutron stars may well be in the useful range 
$\Gamma=10^{10}-10^{11}$. In order to escape freely from the plerion these 
nuclei should be able to cross the ejecta without suffering spallation and 
photodisintegration, which results in a constraint on the magnetic field of 
the neutron star and on its rotation period \cite{boe}. In case of Iron nuclei,
the origin may be galactic, although this possibility does not appear to be 
favored by current data, due to the lack of evidence for anisotropy connected 
to the galactic disk. In other neutron star based models, where the accelerated
particles are protons \cite{lazarian,arons}, the sources are localized in other
galaxies, and therefore the spectrum of UHECRs at the Earth has the usual 
GZK feature. 

\subsection{Gamma Ray Bursts}

Cosmic rays may be accelerated diffusively at the relativistic shock front 
created by the relativistic fireball of a Gamma Ray Burst (GRB) 
\cite{vietri95,wax95}. 
When the burst explodes in the interstellar medium, with magnetic field in 
the $\mu G$ range, the maximum energy of the accelerated particles is 
\cite{gallant99} $E_{max}\approx 10^{15} eV B_\mu$, where $B_\mu$ is the 
magnetic field in $\mu G$. On the other hand, if for instance the GRB 
goes off in the relativistic wind of a neutron star 
\cite{narayan92,VietriStella}, then the magnetic field is expected to be larger
and the maximum energy can be in the $10^{20}$ eV range \cite{vietri2003}.
Acceleration of UHECRs in GRBs may also occur due to other plasma physics 
inspired processes, such as the wakefield acceleration proposed in 
\cite{pisin}.

If UHECRs are accelerated in GRBs, observations require the presence of an 
intergalactic magnetic field, in order to dilute in time the arrival of 
charged particles generated in the few GRB events occurring within $\sim 100$ 
Mpc. Depending on the power spectrum associated with such field, the upper 
limits derived from Faraday rotation measurement are in the range
$10^{-9}-10^{-11}$ Gauss \cite{burles}. In a magnetic field of this order of 
magnitude, the average deflection angle of UHECRs is smaller than the angular 
resolution of present experiments (2-3 degrees), therefore clusters of events 
are expected. If the observed multiplets are in fact the result of bursting 
sources, the higher energy particles should always reach the detector earlier
than the lower energy ones. Although this condition is not satisfied by 
the AGASA clustered events, it was proposed \cite{vietri2003} that 
fluctuations may in fact invert the order of arrival of particles with energy.
The probability to obtain the observed multiplets simply due to statistical 
fluctuations has however not been calculated yet. 

The energetic input of GRBs in the form of UHECRs was first investigated in 
\cite{stecker2000,scully2002,bere2002}. If GRBs follow approximately the 
star-formation history, then the local rate of bursts should be unable to 
account for the observed flux of cosmic rays with energy above $10^{20}$ eV. 
The arguments presented in \cite{stecker2000} were recently addressed in Ref.
\cite{vietri2003}, where the burst rate and energy depositions were taken from 
recent literature \cite{schmidt2001,frail2001}, and the spectra of UHECRs were 
calculated adopting a red-shift luminosity evolution of GRBs that follows 
Ref. \cite{porciani}. The calculated energy injection rate found in 
\cite{vietri2003} through comparison with the HiRes data, 
for an injection spectrum $E^{-2.2}$, is a factor $\sim 6$ larger than that 
available in gamma rays from GRBs (a factor $9$ larger if the AGASA data are 
used), which may represent a serious problem for the GRB models requiring 
high radiative efficiency. 

\section{The top-down alternative}\label{sec:td}

\begin{figure}[ht]
\centerline{\epsfxsize=2.6in\epsfbox{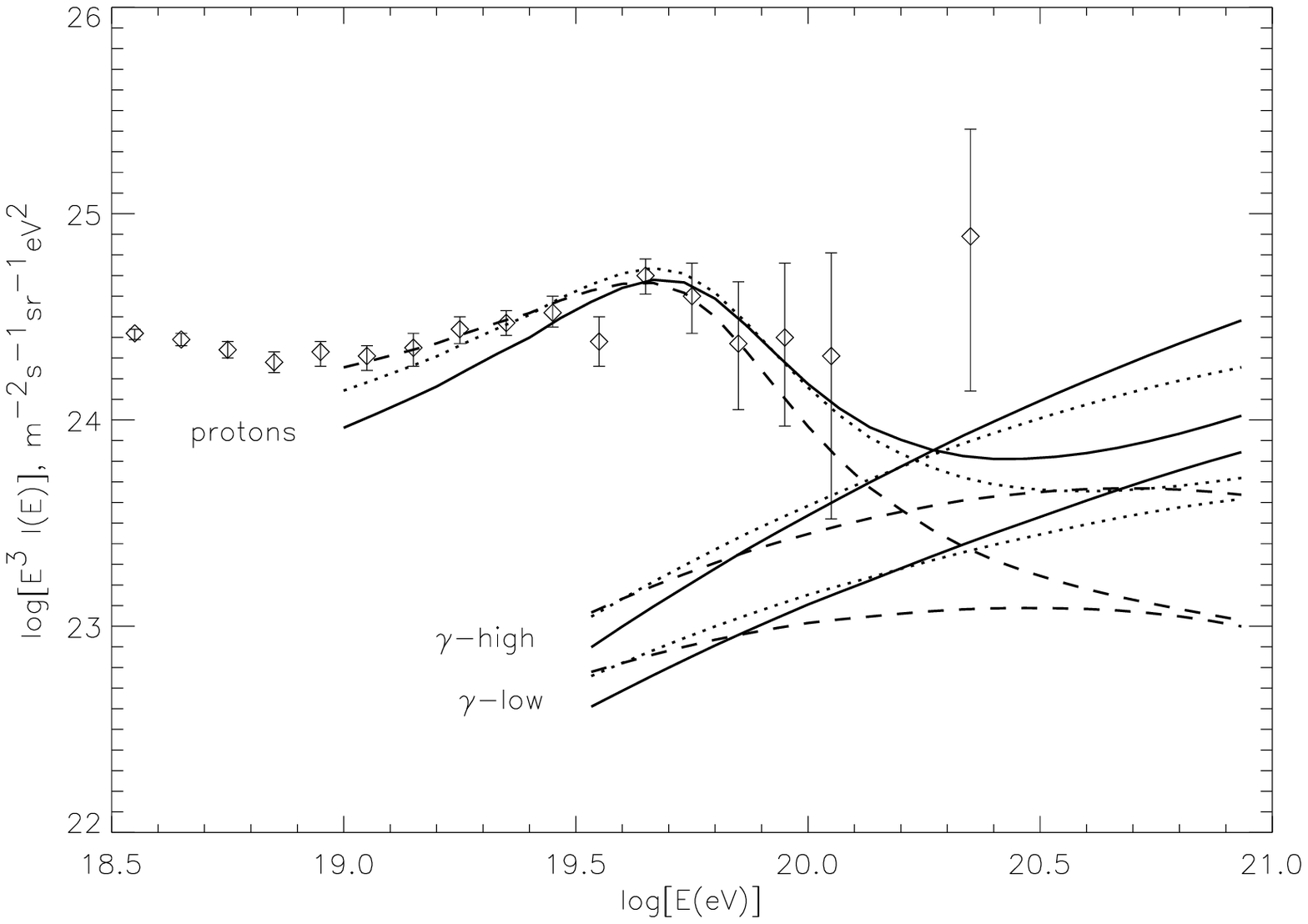}
\epsfxsize=2.6in\epsfbox{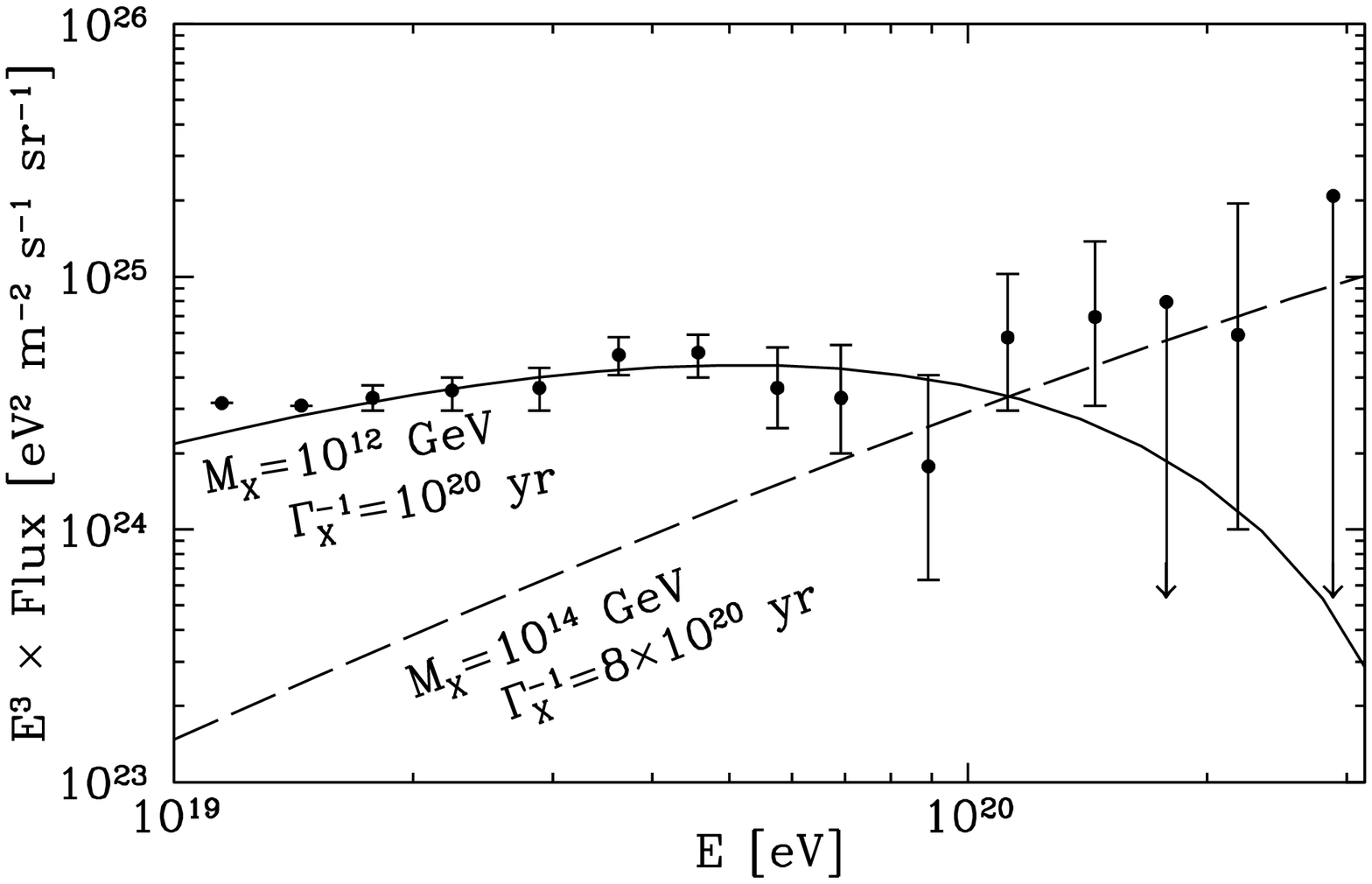}}
\caption{{\it Left}) Spectra of protons and gamma rays from necklaces 
$^{40}$; {\it right}) Spectra of gamma rays from decay of super massive 
relic particles in the halo $^{42}$.}
\label{fig:neck}
\end{figure}

The idea that cosmic rays with extremely high energies may arise from 
the decay of supermassive unstable particles was proposed in the seminal
paper of Ref. \cite{hsw87}. More recently it has become clear that these
supermassive particles may be either the result of the decay or annihilation
of topological defects or quasi-stable relics 
of the inflationary era \cite{berekacvil97,kr98,ckr98,kt98,ktrev}. 
The basic idea, common to all top-down models, is that the decay of a 
supermassive particle results in the production of a quark-antiquark pair 
that hadronizes into mesons and protons. At the source, the composition of 
the produced particles is dominated by gamma rays and neutrinos, while only 
about $5\%$ of the energy goes into protons. After propagation over 
cosmological distances, the relative abundance of protons and gamma rays is 
such that protons dominate up to energies in excess of $10^{20}$ eV, while 
at even higher energies, the composition is expected to be gamma ray dominated 
(see Fig. \ref{fig:neck}). This is the case for most topological defects 
models. In these models, the gamma ray flux is in fact uncertain mainly as
a consequence of the difficulty in measuring the low frequency radio background
from within our galaxy (we are screened by free-free absorption). In Fig.
\ref{fig:neck} the two sets of curves labelled as {\it $\gamma$-high} and
{\it $\gamma$-low} should bracket the range of uncertainty in the radio 
background at the relevant frequencies.
Much discussion exists on which topological defects may generate the observed
fluxes, as summarized in \cite{bbv,bhattasigl}. 

Supermassive particles can also be produced in the early universe independently
of topological defects, either through {\it gravitational production} 
\cite{zelsta72} or through direct coupling to the inflaton field. Relic 
particles produced gravitationally have mass $m_X\leq H(t)\leq  m_\phi$, 
where $H(t)$ is the Hubble constant and $m_\phi$ is the inflaton mass; if 
their decay time is sufficiently long, they can be natural candidates for 
cold dark matter \cite{ckr98,kt98}. The long decay times may be the result 
of some weakly broken discrete symmetry (such as R-parity for the case of 
neutralinos). Supermassive relics accumulate in dark matter halos, and in 
particular in the halo of the Milky Way. Their rare decays may be sufficient 
to account for the observed fluxes of UHECRs \underline{above $10^{20}$ eV} 
\cite{berekacvil97}. In this case the composition is expected to be dominated 
by gamma rays: this is not in contradiction with the findings of \cite{HP}, 
that are stringent for lower energies (the constraint may be important for 
low values of the mass, as for instance for the case plotted as a solid line 
in the right panel of Fig. \ref{fig:neck}). 
The strongest signature of the model is the anisotropy due to the asymmetric 
position of the sun in the Galaxy \cite{dubo98,bbv,beremika99,medwat99}. 
Current observations are not yet stringent enough to rule against or in favor
of the model.

\section{Discussion and conclusions}\label{sec:concl}

\begin{figure}[ht]
\centerline{\epsfxsize=2.5in\epsfbox{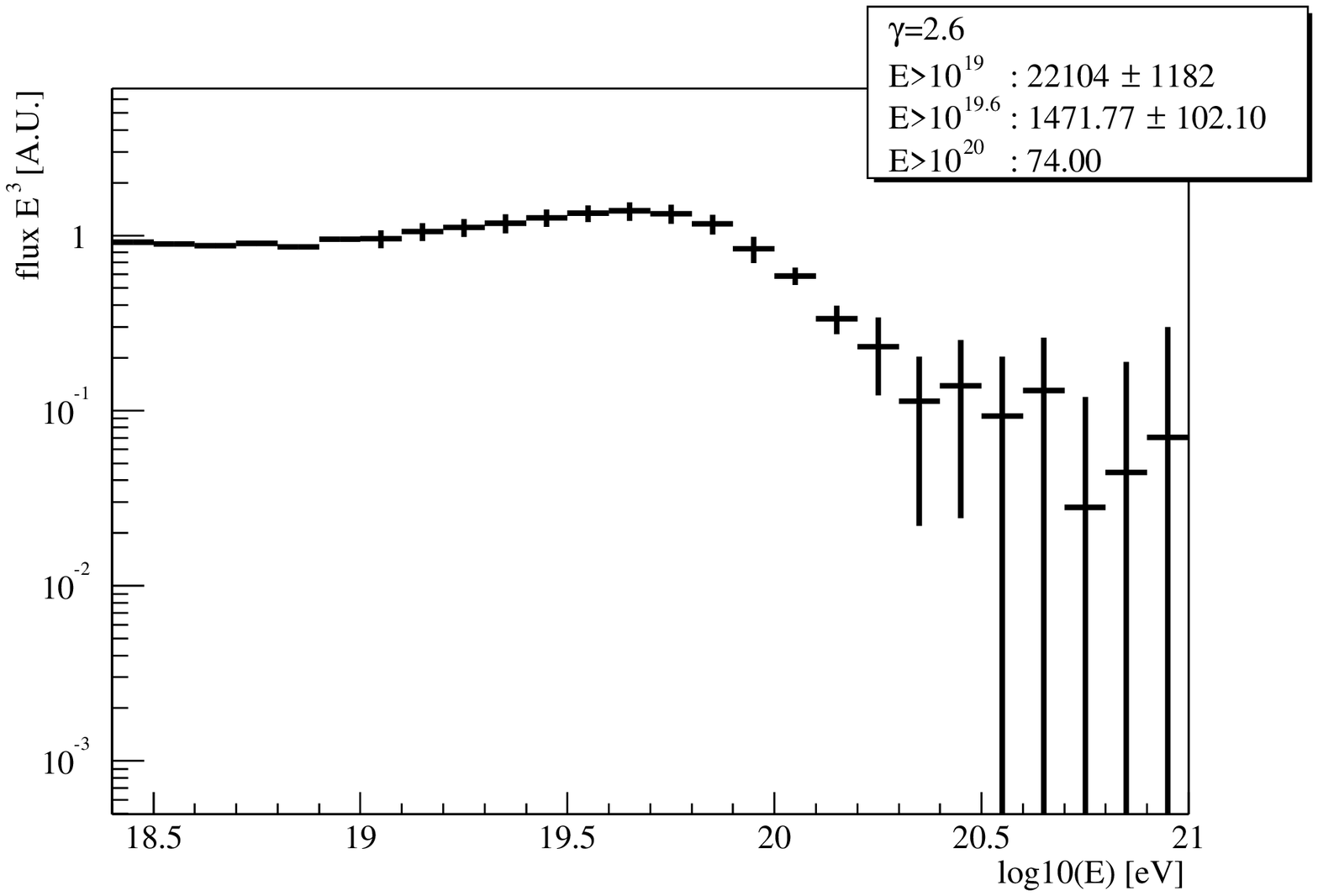}
\epsfxsize=2.5in\epsfbox{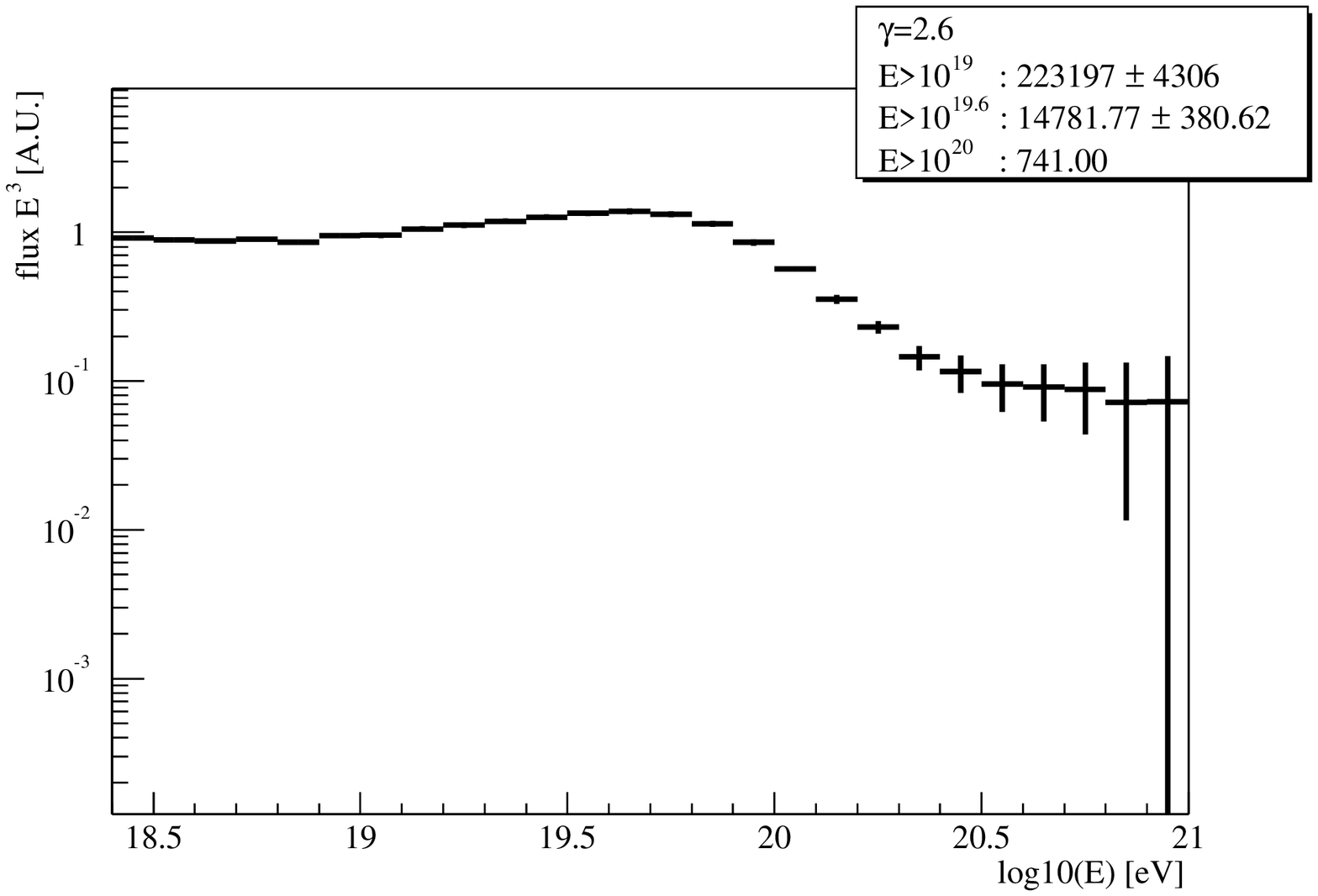}}
\caption{Expected simulated performances of Auger (left panel) and EUSO (right
panel)$^4$.}
\label{fig:future}
\end{figure}

The spectrum of UHECRs and in particular the presence of a GZK feature 
will be properly measured by the next generation cosmic ray experiments,
namely the Auger project \cite{cronin}, and the space-borne EUSO observatory
\cite{scarsi}. The increase in the statistics that will be achieved by these 
two observational enterprises is well illustrated in Fig. \ref{fig:future}, 
where the simulations in \cite{us} have been used in order to predict the 
error bars expected after three years of operation of Auger and EUSO. 
The question about the origin of UHECRs will be answered when the 
spectrum in the energy region above few $10^{19}$ eV will be reliably 
measured: if the GZK feature is there, then we will have a proof that 
UHECRs are extragalactic protons generated by astrophysical sources. 
The high statistics of large multiplicity clusters of events will on
the other hand give a direct hint on the type of sources that we have to 
look for, because the combination of the spectrum and small scale 
anisotropies provides a potentially powerful tool to determine not only
the energy injected per unit volume but also the energy per source.
On the other hand, if no evidence for the GZK feature will be found, 
then it will probably be hard to avoid to invoke some kind on new physics.
In this case the study of composition might be the tool to discriminate
among different models.

\end{document}